# Studies on optical signal due to oxygen effect on hydrogenated amorphous/crystalline silicon thin-films

Meenakshi Rana[1], Chandan Banerjee[2] and Papia Chowdhury[1]*

[1]Department of Physics and Materials Science & Engineering, Jaypee Institute of Information Technology, Noida 201307, Uttar Pradesh, India

[2]National Institute of Solar Energy, Gwalpahari, Gurugram, Haryana 122003, India

## Abstract

We have studied the effects of oxygen on hydrogenated amorphous/crystalline silicon films in terms of their structural and optical properties. Different "hydrogenated silicon oxide" (SiO:H) and "silicon" (Si:H) films are fabricated between microcrystalline and amorphous transition region. X-ray diffraction, Raman, FTIR and UV-Vis emission spectrometry have been used to characterize different films. A comparison of the results with those of different types of films like "hydrogenated amorphous silicon oxide" (a-SiO:H), "hydrogenated amorphous silicon" (a-Si:H) and "microcrystalline silicon" (μc-Si:H) films reveal their superiority as an excellent substance for solar cell. X-ray diffraction, FTIR and Raman spectral analysis show that difference of the H dilution effect has a major effect on the structure of the film and the optical properties. Photoluminescence analysis of amorphous silicon–oxygen and silicon-hydride alloy films has established their efficient application appropriate as Si based light emitting devices. A large optical band gap of 1.83 eV and appearance of strong photo luminescence at 2.0 eV validates the applicability of a-SiO:H film as a better alternative for the solar cells.

* Corresponding author, Fax: +91 120 2400986

E-mail: papia.chowdhury@jiit.ac.in

# 1. Introduction:

Amorphous silicon is a very well known base material for higher conductivity. Out of different noncrystalline semiconductor materials like: selenium, etc. though they have very good electronic properties, they are proved unsuitable for making of efficient solar cells [1]. After the work of Chittick et.al [2] it was proved that as crystalline silicon, amorphous silicon can also have a high conductivity in the presence of some doping. The first industrial solar cell was reported by doped amorphous silicon on 1976 by Christopher Wronski [3]. Normally, hydrogen atoms are found in amorphous silicon in a considerable fraction. These hyderogen atoms play an important role for the enhancement in the electronic properties of amorphous silicon structure. It is generally known as "hydrogenated amorphous silicon (a-Si:H)"[4]. Photo voltaic (PV) technology based on a-Si:H is exceptional compared with other available PV technologies [5]. The importance of a-Si:H comes into the picture due to its promising optical properties for collecting solar energy. The technology based on a-Si:H is also very easy and low-cost in comparison of other available technologies for making crystals such as crystalline Si (c-Si) [6]. Due to the amorphous nature of a-Si, it absorbs sunlight extra powerfully than other crystalline and poly crystalline silicon (c-Si and poly-Si). The selection rules, which weaken the absorption in c-Si an indirect band gap type of semiconductor, which is not applicable to a-Si. "Band gap" of a-Si is considerably larger than c-Si, which reduces optical absorption of c-Si than a-Si [7,8] of same thickness. To absorb the same energy, a c-Si layer needs to be much thicker. This indicates that solar cell composed from a-Si required less material than from c-Si. a-Si can be made at a very low temperature on inexpensive substrate like glass and so the product can be made through a low-cost process. The thickness of the absorbing layers in a-Si solar cells is less than 1 micron. Another important quality of a-Si selected substrates is that it can be made light weight and flexible. These effects are very significant for numerous applications. A positive temperature coefficient is also observe for output power of a-Si PV products. In area having extra sunshine, a-Si based devices observed to posses higher efficiency at higher temperature. The conductivity of a-Si can be modified by the effect of p type or n type doping on it. For n type doping conductivity can be induced due to mobile electrons. For p type doping, the induced conductivity appears due to mobile wholes. The energy payback time is also observed to be very less in a-Si products as compared to c-Si [9]. For thin film silicon solar cell, silicon usually is deposited on a glass substrate. There are lots of research works have been reported on hydrogenated μc-Si:H thin film solar cells [10-12] till now, research works on a-Si solar cells are very few. Out of differently available hydrogenated amorphous phase silicon, one is hydrogenated amorphous silicon oxide (a-SiO:H), where oxygen is attached with silicon. The SiO:H films have more advantages over Si:H film because of its wide band gap. The reason of the wide band gap is the presence of oxygen. Presence of oxygen increases the band gap of silicon [13]. Oxygen rich phase in SiO:H films helpful to increase band gap whereas the silicon rich phase in SiO:H films contributes towards higher conductivity [14]. Many research works have been reported on optical (optical absorption coefficient, integrated solar irradiance) and electrical (band gap, voltage, efficiency) properties of hydrogenated amorphous silicon solar cells which get affected by its oxygen content [15,16] in the form of a-SiO:H. But very few information about the properties of a-SiO:H has been explored till now.

Generally, Plasma enhanced chemical vapor deposition (PECVD) system is used to deposit hydrogenated silicon thin films by glowing discharge decomposition of silane [17,18]. This silane gas is diluted with hydrogen and have many significant impact on the characteristics of a-Si:H films [19,20]. By changing the deposition parameters such as; chamber pressure, gas mixture composition, flow rates, RF power density and substrate temperature, one can change structure of the films. Depending on the deposition parameters, amorphous, micro/nano crystalline and procrystalline [21] structures of hydrogenated silicon (Si:H) films can be prepared. These different Si:H structures can be used for enhancing the performance of silicon based thin film solar cells.

In the present work we have synthesized and characterized some doped and intrinsic hydrogenated amorphous and microcrystalline SiO:H and Si:H thin films on glass substrate. We have developed some n and p type µc-SiO: H and amorphous thin films having suitable characteristic properties used for the fabrication of amorphous silicon (a-Si) solar cells. We will try to characterize these films to describe some important aspects of their electrical and optical properties in view of their probable application as solar cells. We have also reported the results in view of comparison of SiO:H with Si:H film.

## 2. Materials and Methods
## 2.1 Materials:

Different types of silicon-based materials used in the present study have been prepared by using "five chamber cluster tool plasma enhanced chemical vapour deposition (PECVD) unit" [22]. The same chamber has been used to grow n-doped a-Si:H and µc-Si:H. PH$_3$ (1%) in SiH$_4$ and H$_2$ gases were used to grow a-Si:H and μc-Si:H films. To grow μc-SiO:H films CO$_2$ gas was employed [22].

## 2.1 Experimental Methods:

For measuring X-ray diffraction (XRD) patterns, "Bruker D8 Advance diffractometer has CuKα" as radiation source have been used. Raman spectra have been measured with "Raman spectrometer equipped with a SPEX TRIAX 550 monochromator (model GDLM-5015 L, 8mw)" and a "liquid-nitrogen-cooled charge-coupled device (CCD; Spectrum One with CCD 3000 controller)". Spectral acquisition time and spectral resolution was 1 min and 2 cm$^{-1}$ respectively for the Raman system. FTIR spectra were recorded with a "Perkin-Elmer FTIR spectrophotometer (model Spectrum BX-II source: nichrome glower wire with DTGS detector)". The spectral resolution of IR spectrometer was 2 cm$^{-1}$ with a range 400−4000 cm$^{-1}$. UV-Vis absorption spectra were measured with a "Perkin-Elmer absorption spectrophotometer (model Lamda-35, source: tungsten iodide and deuterium)" in the 200−1100 nm spectral range with a 0.5 and 1.0 nm slit width. The UV-Vis fluorescence spectra were recorded with "Perkin-Elmer luminescence spectrophotometer (model Fluorescence-55 source: xenon)" in the 200−900 nm spectral range.

## 3. Results and Discussion:

We have discussed the properties of a-SiO:H and a-Si:H films, that are placed on the top of glass substrate. Figure 1 shows the XRD spectra of these films as compared with the XRD spectra of crystalline Si wafer and glass. Normally, Si wafer shows five distinct peaks at 21°, 39.3°, 43°, 58° and 69.4° correspond to ⟨111⟩, ⟨200⟩, ⟨220⟩, ⟨311⟩ and ⟨400⟩ planes of crystalline silicon phases [23]. Whereas glass shows only a wide, band around 20.8° due to its amorphous phase [24].

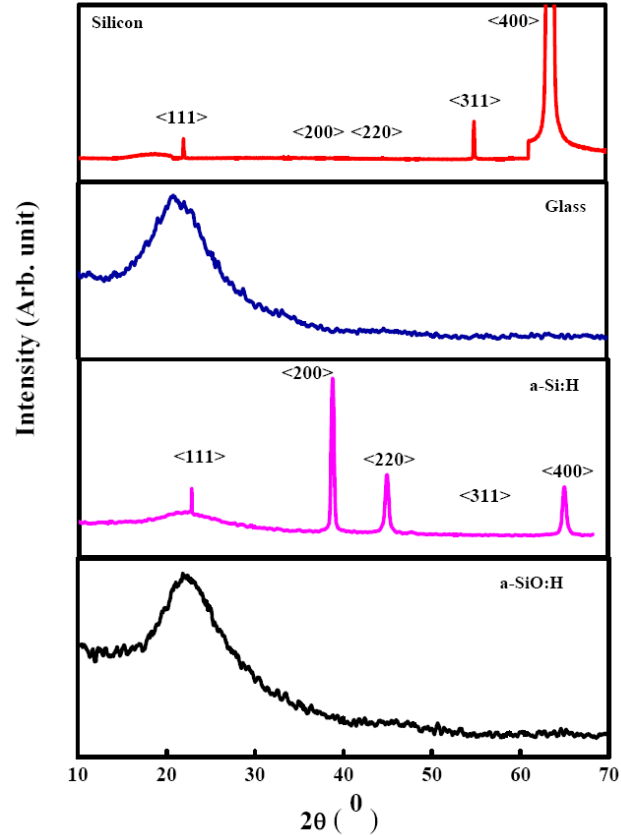

**Figure 1:** XRD pattern of crystalline silicon, glass, amorphous a-SiO:H and a-Si:H films placed on top of glass.

In the presence of different a-Si films, both the samples show different characteristics. In a-Si:H film in addition to all silicon crystalline phases a wide peak at 22.1° appears due to the amorphous phase of silica whereas in a-SiO:H film due to excess oxygen content the crystalline nature of Si disappears totally and amorphous nature dominates with the appearance of only one wide band at 22.2°.

Raman spectra of different doped crystalline and amorphous SiO:H and Si:H films have been taken (Figure 2). A broadband centred at 490 cm$^{-1}$ in SiO:H films depicted the amorphous phase of it. Shift from a-SiO:H to μc-Si growth shown by an emergence of thin and sharp band at 528 cm$^{-1}$. When we compare Raman intensities of the SiO:H films with Si:H films, we observe that with increasing H ratio the comparative intensity of the crystalline peak increases. This clearly indicates the good conformity with the results obtained from x-ray diffraction pattern. The whole Raman spectrum divided into three components as crystalline component, amorphous component and intermediate component. The peak is observed at 528 cm$^{-1}$ due to crystalline component, peak at 485 cm$^{-1}$ due to amorphous component and intermediate component peaked between the ranges 495 – 520 cm$^{-1}$ that is associated with bond dilation at grain boundaries.

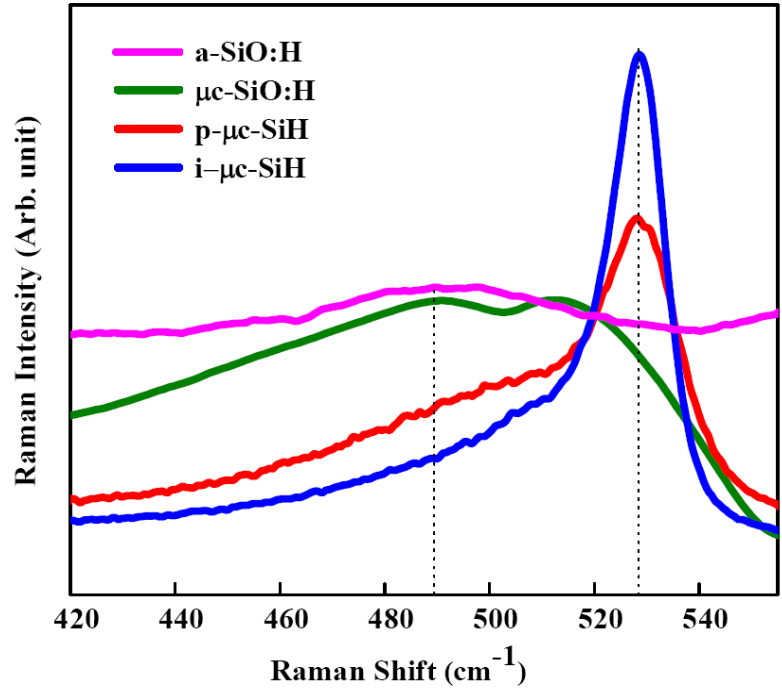

**Figure 2:** Raman spectra of doped "SiO:H" and "Si:H" films placed on the top of the glass substrate

By using Raman spectral results, the crystalline volume fraction ($X_C$) have been deduced as:

$$X_C(\%) = \left[ (I_i + I_C) / (I_C + I_i + I_a) \right] \times 100 \qquad (1)$$

Where $I_c$, is the integrated intensities of the crystalline $I_a$, is the integrated intensities of the amorphous and $I_i$ is the integrated intensities of the intermediate peaks. The observed result shows that $X_C$ gradually increases from amorphous to microcrystalline SiO:H films. Also $X_C$ increases monotonically from 63.5% to 89.3% from μc-SiO:H to μc-Si:H films as the H dilution increases to the μc- Si growth which reflects with the narrow band appearance at 528 cm$^{-1}$.

Similarly, the grain boundary volume fraction $X_i$ has been calculated by the equation [25]

$$X_i(\%) = \left[ (I_i) / (I_C + I_i + I_a) \right] \times 100 \qquad (2)$$

It is found that as H dilution increases the $X_i$ decreases from 35.2 % to 29.1%. So an opposite behaviour is observed in grain size and crystalline volume fraction with increase the phase from amorphous to crystalline as well as with increasing H dilution.

The bonded hydrogen content and the effects of oxygen atom have been also studied from the FTIR absorption spectra as shown in figure 3. Structural characterization of amorphous films is extremely difficult. According to the "random bonding model (RBM)" [26], "unhydrogenated silicon oxide films ($SiO_x$) are composed of five basic bonding configurations", Si ($Si_{4-n}O_n$), n=0-4 with basic bonding units as: $Si_2O$, SiO, $Si_2O_3$, $SiO_2$.

According to modified RBM, in hydrogenated films (a-$SiO_x$:H) the incorporation of Si-H bonds can be performed by replacing an Si nearest neighbour to the Si site by an H atom in Si. In hydrogenated Si film, Si-H stretching absorption peaks correspond to Si-H stretching is observed at 2337 cm$^{-1}$ and 2365 cm$^{-1}$. Due to the attachment of oxygen atom, the peak get shifted to a higher wave number side associated with H-Si ($Si_{3-n}O_n$) for n = 0, 1, 2, 3 respectively with an appearance of new broad peak at 2000 cm$^{-1}$. Therefore, the Si-H stretching band can be dissolved into three components at 2000, 2337 and 2365 cm$^{-1}$ due to H-Si ($OSi_2$), H-Si ($O_2Si$), H-Si ($O_3$). Again a new band appears in a-Slosh films at 1386 cm$^{-1}$ due to Si-O-Si symmetric stretching vibration which is absent in hydrogenated silicon.

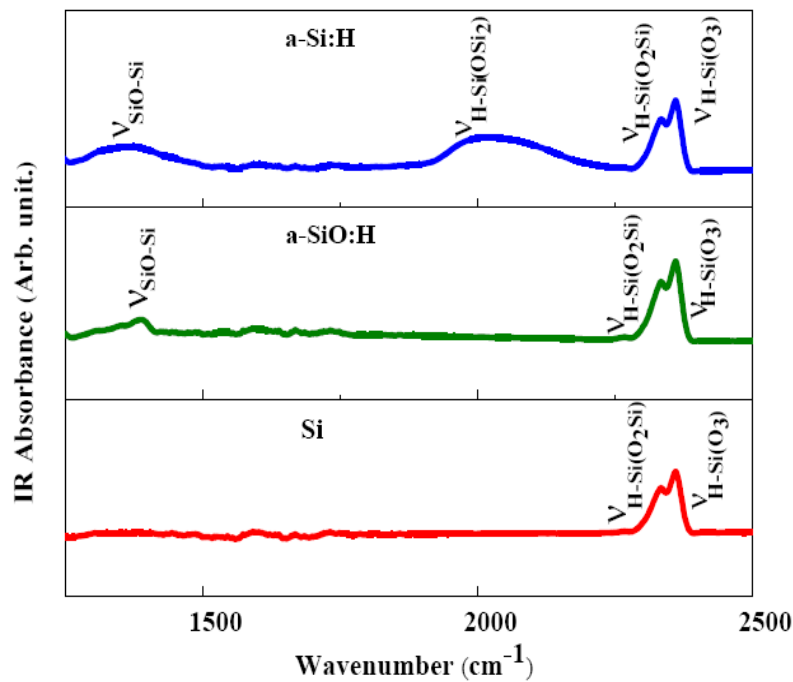

**Figure 3:** FTIR spectra of hydrogenated silicon and SiO:H film deposited on a glass substrate.

However, the crystalline and grain size, volume fraction behaves differently with the increment of H dilution. However, it is not obvious that how the optical and electronic properties of the films will change with changing the structure of the films from SiO:H to Si:H. We have carried out UV-Vis absorption and luminescence studies. Figure 4 shows the absorption spectra of both the films.

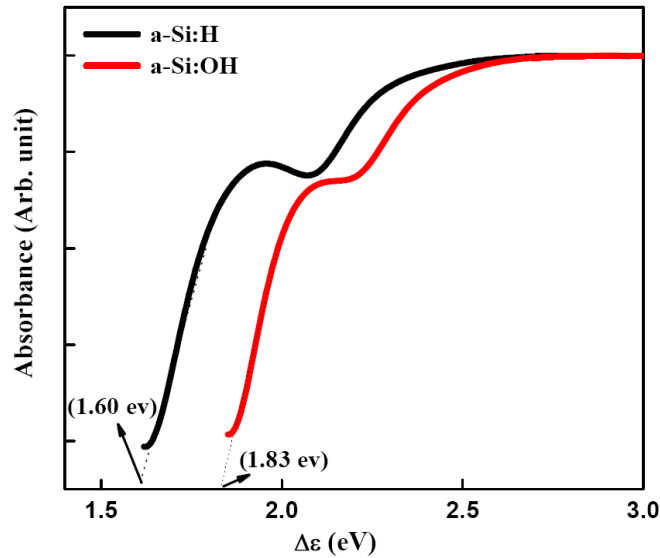

**Figure 4:** UV-Vis absorption spectra of a-SiO:H and a-Si:H films.

To calculate the optical band gap, we have been used the tauc model. It uses experimental spectroscopic absorbance from the figure 4 we can estimate that band gap ($\Delta\varepsilon$) of a-SiO:H film comes out to be 1.83 eV whereas $\Delta\varepsilon$ for a-Si:H observed as 1.6 eV which perfectly matches with the conventional literature [27] values. According to the Watanabe model [28] SiO:H belongs to a two phase material having an island of SiO in a matrix of a-Si:H. The two-phase model suggested that the oxygen rich phase is efficient in optical band gap increment. This is one of the advantages that why a-SiO:H films are widely used as the window layer of the solar cells in research and development level.

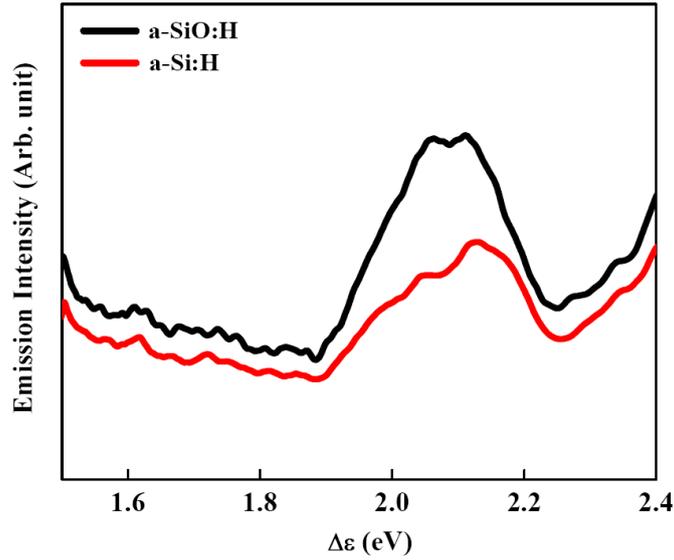

**Figure 5:** Emission spectra of "a-SiO:H" and "a-Si:H" films.

SiO:H and Si:H films have tunable optical band gap. The red, blue coloured emission and their combination mechanism from silicon based materials have been studied by different researchers [29-33]. Kanashima et al. [34] have been reported dual emission at 3.5 eV (354.24 nm) and 4 eV (309.96 nm) from $SiO_2$ films. But there are very few studies on "a-SiO:H" and "a-Si:H" films. Figure 5 shows different a-SiO:H and a-Si:H films photoluminescence spectra with excitation at 3.68 eV (336.91 nm) at room temperature. a-SiO:H film shows two strong emission peaks at 2.05 eV (604.80 nm) and 2.10 eV (590.40 nm) with two weak shoulders at higher energies at 2.34 eV (529.84 nm) and one at 1.60 eV (774.90 nm) lower energy side respectively. Whereas in a-Si:H film the emission band at 2.05 eV (604.80 nm) disappears with the existence of only one strong emission at 2.09 eV (585 nm).

| Nature of film | Raman shift ($\bar{\upsilon}$) in ($cm^{-1}$) | | Volume fraction from Raman (%) | | Optical band gap ($\Delta\varepsilon$) in (eV) |
|---|---|---|---|---|---|
| | | | $X_c$ | $X_i$ | |
| μc-Si:H | - | 528 | 89.3 | 29.11 | |
| a-Si:H | | | | | 1.60 |
| μc –SiO:H | 490 | 510 | 70 | 35.21 | |
| a-SiO:H | 490 | - | 63.5 | 32.27 | 1.83 |

**Table: 1** Comparison of Raman, UV-Vis data for SiO:H and Si:H films.

The two remaining shoulders exists at the same positions as that of a-SiO:H film at higher energies at 2.34 eV (529.84 nm), and 1.60 eV (774.90 nm) respectively. Due to the presence of oxygen content in a-SiO:H film, a red shift in the emission peak at 2.10 eV (590.40 nm) is observed. So the luminescence (visible) properties associated with excitation (3.68 eV or 336.91 nm) from a-SiO:H and a-Si:H films differ with the available oxygen content.

## 4. Conclusion:

We have studied the property of oxygen on the structural and optical properties of the different doped and undoped hydrogenated a-SiO:H and a-Si:H films. This work systematically study of amorphous Si and microcrystalline Si structure of films based on the hydrogen dilution and oxygen incorporation effects on them. Different XRD, Raman and FTIR spectral analysis display that difference of the H dilution effect has a major effect on the structure of the film and the optical properties. A structural change from a-Si to µc-Si is caused by the increasing H ratio. This structural change creates a regular alter of the electronic structures which is observed by the UV-Vis absorption and photoluminescence spectroscopy. We have also observe a very remarkable type of the high energy ($cm^{-1}$) Raman from µc-Si phase to a-Si phase both in SiO:H and Si:H. Major optical change we observed of higher $cm^{-1}$ Raman from µc-Si phase such as energy blue shift and decrement in the intensity and increment in the band width to amorphous phase. The same observation has been verified by the FTIR data. The observable photoluminescence (visible) properties with excitation (3.68 eV) from different hydrogenated Si films changes with oxygen content. The decrease in 2.05 eV emission band intensity observed in a-Si:H is tentatively due to the absence of oxygen. In conclusion, combining XRD, UV-Vis, Raman and IR spectra give us a rich information about the SiO:H and Si:H films which are very crucial to know different optoelectronic properties of µc-Si and a-Si films.

## References:

[1] Tomar N, Agrawal A, Dhaka VS, Surolia PK (2020) Ruthenium complexes based dye sensitized solar cells: Fundamentals and research trends. Solar Energy, 207:59-76.
[2] Chittick RC, Alexander JH, Sterling HF (1969) The Preparation and Properties of Amorphous Silicon. J. Electrochem. Soc 116:77-81.
[3] Carlson DE, Wronski CR (1976) Amorphous Silicon Solar Cells, Appl. Phys. Lett. 28: 671-673.
[4] Schiff EA, Hegedus S, Deng X (2011) Amorphous silicon-based solar cells. Handbook of photovoltaic science and engineering, 487-545.
[5] Ramanujam J, Bishop DM, Todorov TK, Gunawan O, Rath J, Nekovei R, Romeo A (2020) Flexible CIGS, CdTe and a-Si: H based thin film solar cells: A review. Progress in Materials Science, 110:100619.
[6] Chopra KL, Paulson PD, Dutta V (2004) Thin-film solar cells: an overview. Progress in Photovoltaics: Research and applications. 12: 69-92.
[7] Vaněček MA, Poruba A, Remeš Z, Beck N, Nesládek M (1998) Optical properties of microcrystalline materials. J. Non-Cryst. Solids 967: 227-230.
[8] Sriprapha K, Sitthiphol N, Sangkhawong P, Sangsuwan V, Limmanee A, Sritharathikhun J (2011) p-Type hydrogenated silicon oxide thin film deposited near amorphous to microcrystalline phase transition and its application to solar cells. Current Applied Physics, 11: S47-S49.
[9] Hagedorn G (1989) Hidden energy in solar cells and photovoltaic power stations. Proceedings of the 9$^{th}$ European Solar Energy Conference, Freiburg, 542–545.
[10] Zhao Y, Zhang X, Bai L, Yan B (2019) Hydrogenated Microcrystalline Silicon Thin Films. Handbook of Photovoltaic Silicon, 693:756.
[11] Tabuchi T, Toyoshima Y, Fujimoto S, Takashiri M (2020) Optimized hydrogen concentration within a remotely induced hollow-anode plasma for fast chemical-vapor-deposition of photosensitive and< 110>-preferential microcrystalline silicon thin-films. Thin Solid Films, 694: 137714.
[12] Mandal S, Das G, Dhar S, Tomy RM, Mukhopadhyay S, Banerjee C, Barua AK (2015) Development of a novel fluorinated n-nc-SiO: H material for solar cell application. Materials Chemistry and Physics, 157: 130-137.
[13] Ma HP, Lu HL, Yang JH, Li XX, Wang T, Huang W, Zhang DW (2018) Measurements of microstructural, chemical, optical, and electrical properties of silicon-oxygen-nitrogen films prepared by plasma-enhanced atomic layer deposition. Nanomaterials, 8: 1008.
[14] Das D, Iftiquar SM, Barua AK (1997) Wide optical-gap a-SiO: H films prepared by rf glow discharge. Journal of non-crystalline solids. 210: 148-154.
[15] Zhu M, Han Y, Godet C, Wehrspohn RB (1999) Photoluminescence from hydrogenated amorphous silicon oxide thin films. Journal of Non-Crystalline Solids. 254: 74-79.
[16] Ogale AS, Ogale SB, Ramesh R, Venkatesan T (1999) Octahedral cation site disorder effects on magnetization in double-perovskite Sr 2 FeMoO 6: Monte Carlo simulation study. Applied physics letters.
[17] Chen Y, Wang J, Lu J, Zheng W, Gu J, Yang S, Gao X (2008). Microcrystalline silicon grown by VHF PECVD and the fabrication of solar cells. Sol. Energy. 82: 1083–1087,
[18] Klein S, Repmann T, Brammer T (2004) Microcrystalline silicon films and solar cells deposited by PECVD and HWCVD. Sol. Energy, 77:893–908.

[19] Kroll U, Meier J, Shah A, Mikhailov S, Weber J (1996) Hydrogen in amorphous and microcrystalline silicon films prepared by hydrogen dilution. J. Appl. Phys., 80: 4971.

[20] Yang J, Xu X, Guha S (1994) Stability studies of hydrogenated amorphous silicon alloy solar cells prepared with hydrogen dilution. In: Symposium A – Amorphous Silicon Technology – 1994, MRS Online Proceedings Library, 33:687.

[21] Asano A, (1990) Effects of hydrogen atoms on the network structure of hydrogenated amorphous and microcrystalline silicon thin films. Appl. Phys. Lett. 56: 533.

[22] Das G, Bose S, Mukhopadhyay S, Banerjee C, Barua AK (2019) Innovative Utilization of Improved n-doped μc-SiO x: H Films to Amplify the Performance of Micromorph Solar Cells. Silicon, 11: 487-493.

[23] Vargas-Ortiz RA, Espinoza-Beltran FJ (2005) Structural and chemical composition of si-al oxy-nitride coatings produced by reactive dc magnetron sputtering. Advances in Technology of Materials and Materials Processing Journal. 7: 87-90.

[24] Polosan S, Galca AC, Secu M (2011) Band-gap correlations in $Bi_4Ge_3O_{12}$ amorphous and glass ceramic materials. Solid State Sciences. 13: 49-53.

[25] Zhao H, De Geuser, F, da Silva AK, Szczepaniak A, Gault B, Ponge D, Raabe D (2018) Segregation assisted grain boundary precipitation in a model Al-Zn-Mg-Cu alloy. Acta Materialia, 156:318-329.

[26] Philipp HR (1972) Optical and bonding model for non-crystalline SiOx and SiOxNy materials. Journal of Non-Crystalline Solids 8:627-632.

[27] Borchert D, Grabosch G, Fahrner WR (1997) Preparation of (n) a-Si: H/(p) c-Si heterojunction solar cells. Solar Energy Materials and Solar Cells, 49: 53-59.

[28] Watanabe H, Haga K, Lohner T (1993) Structure of high-photosensitivity silicon-oxygen alloy films. Journal of non-crystalline solids. 164: 1085-1088.

[29] Canham LT (1990) Silicon quantum wire array fabrication by electrochemical and chemical dissolution of wafers. Appl. Phys. Lett. 57:1046.

[30] Kenyon AJ, Trwoga PF, Pitt CW, Rehm G (1996) The origin of photoluminescence from thin films of silicon-rich silica. Journal of Applied Physics. 79: 9291-9300.

[31] Zhao X, Schoenfeld O, Komuro S, Aoyagi Y, Sugano T (1994) Quantum confinement in nanometer-sized silicon crystallites. Physical Review B, 50:18654.

[32] Shen H, Gao Q, Zhang Y, Lin Y, Lin Q, Li Z, Wang S (2019). Visible quantum dot light-emitting diodes with simultaneous high brightness and efficiency. Nature Photonics, 13:192-197.

[33] Xu K, Chen Y, Okhai TA, Snyman LW (2019). Micro optical sensors based on avalanching silicon light-emitting devices monolithically integrated on chips. Optical Materials Express, 9:3985-3997.

[34] Kanashima T, Okuyama M, Hamakawa Y (1994) Photoluminescence of $SiO_2$ films grown by photo-induced chemical vapor deposition. Applied surface science, 79: 321-326.